\documentclass[pre]{revtex4}

\usepackage{graphicx}
\usepackage{amsmath}
\usepackage{bm}

\begin{document}

\title[Measurement of two-sphere hydrodynamics]{Measurement of the hydrodynamic forces between two polymer-coated spheres}

\author{P. Bartlett}
\author{S. I. Henderson}
\author{S. J. Mitchell}

\affiliation{School of Chemistry, University of Bristol, Bristol
BS8 1TS, UK}

\label{firstpage}

\begin{abstract}

The hydrodynamic forces between Brownian spheres are determined
from a measurement of the correlated thermal fluctuations in
particle position using a new method, two-particle
cross-correlation spectroscopy (TCS). A pair of 1.3 $\mu$m
diameter polymer-coated poly(methyl methacrylate) were held at
separations of between 2$\mu$m and 20$\mu$m using optical traps.
The mobility tensor is determined directly from the
statistically-averaged Brownian fluctuations of the two spheres.
The observed distance dependence of the mobility tensor is in
quantitative agreement with low-Reynolds number calculations.

\end{abstract}

\maketitle
\section{Introduction}

The hydrodynamic interactions between colloidal particles are
important from both a fundamental and an industrial viewpoint.
They determine, for instance, the rheological behaviour of
suspensions, the kinetics of aggregation and phase separation and
many other common colloidal phenomena \cite{Russel-648}. Yet
despite this, the hydrodynamic properties of all but the simplest
colloidal system have been a subject of considerable debate
\cite{Segre-1121}. A key factor in this uncertainty has been the
intrinsically long-ranged nature of the hydrodynamic coupling
between solid particles. In the dilute limit, where it is
sufficient to calculate just the leading order interaction, the
solution of the stationary Stokes equation \cite{1615} reveals
that the interactions decay like the inverse separation (the Oseen
tensor). The non-local nature of such interactions has led to
considerable theoretical and numerical difficulties. Experiments
have,  conversely, also been problematic.  Methods such as light
scattering which have been used extensively in the past to probe
the dynamics of concentration fluctuations provide only limited,
 {\it indirect} information on the microscopic nature of
hydrodynamics in suspensions.

In this paper we report a detailed experimental study of the
distance dependence of  the hydrodynamic interactions between an
individual pair of polymer-coated colloidal particles. Optical
tweezers are used to hold two uncharged poly(methyl methacrylate)
(PMMA) spheres apart at separations from between 2$\mu$m and
20$\mu$m. The hydrodynamic forces are measured using a new, highly
sensitive, experimental technique, two-particle cross-correlation
spectroscopy (TCS). TCS measures the statistically-averaged
instantaneous fluctuations in position of
 two probe spheres. The
in-plane position of each sphere is measured to nanometer
precision using a quadrant photodetector. This technique is
applied to a pair of colloidal particles suspended in a Newtonian
liquid to yield a detailed {\it direct} test of low-Reynolds
number predictions of hydrodynamic forces in a simple colloidal
system.

Several reports of the test of theoretical predictions for the
hydrodynamic coupling between a pair of spheres have already been
published \cite{Crocker-882,Meiners-1501}. Our experiments differ
in several regards. First, in the experiments reported to date,
the surfaces of the colloidal spheres have been bare and not
covered by a polymer layer. Since the adsorption or anchoring of a
polymer onto the surface of a particle is a common method of
imparting colloidal stability it is important to establish if the
presence of a polymer layer modifies the hydrodynamic forces.
Indeed theoretical calculations \cite{1653} predict that, flow
within the polymer layer, removes the divergence of the
hydrodynamic forces seen at small pair-separations. Second, the
reported experiments have used charged spheres at relatively low
electrolyte concentrations (e.g. 0.1mM in \cite{Crocker-882}) so
that residual electrostatic interactions between the charged
spheres or between the confined spheres and the walls complicates
the interpretation. The experiments reported here used an
uncharged colloid, for which previous work
\cite{Pusey-853,Underwood-839} has shown that the interaction
potential is well approximated by that of hard spheres.

Our paper is organised as follows: In the next section we describe
the details of our experiment. Section~\ref{sec:theory} details
the Brownian motion of an isolated sphere in a harmonic optical
potential and describes the motion of a pair of
dynamically-coupled spheres. We present our results in
\S\ref{sec:results} before concluding.

\section{Experimental methods}
\label{sec:exp}

Measurements were performed on a dilute suspension of uncharged
poly(methyl methacrylate) (PMMA) spheres
 of radius 0.65 $\pm$ 0.02$\mu$m. To minimize the van der Waals forces between
 the PMMA spheres
the surface of each sphere was covered with a covalently bound
polymer brush ($\sim 100${\AA} thick) of poly(12-hydroxy stearic
acid)  \cite{Antl-63}. A dilute suspension of the spheres (volume
fraction $\phi \sim 10^{-7}$) in a mixture of cyclohexane and
cis-decalin was confined within a rectangular glass capillary,
170$\mu$m thick. The ends of the capillary were hermetically
sealed with an epoxy resin to prevent evaporation and to minimize
fluid flow. A small concentration of free polymer stabilizer was
added to the suspension to reduce the adsorption of the spheres
onto the glass surfaces of the cell.

A pair of spheres were trapped in a plane about 40$\mu$m above the
lower glass surface of the cell using two optical traps. The traps
were created by focusing orthogonally polarised beams from a
Nd:YAG laser (7910-Y4-106, Spectra Physics) with a wavelength of
$\lambda$ = 1064nm to diffraction-limited spots using an
oil-immersion microscope objective (100x/1.3 NA Plan Neofluar,
Zeiss). The resulting optical gradient forces localise the sphere
near the focus of the beam. The centre-to-centre separation of the
two traps could be varied continuously from between 2 and
30$\mu$m. Accurate positions for the optical traps were determined
by digitising an image of two trapped spheres with a Imaging
Technology MFG-3M-V frame grabber. The spheres' locations were
measured to within 40nm using a centroid tracking algorithm.

While the {\it mean} position of each sphere is fixed by the
position of the corresponding laser beam, fluctuating thermal
forces cause small but continuous displacements of the particle
away from the centre of the trap. For small displacements the
optical trapping potential is accurately described by a harmonic
potential \cite{Tlusty-1543}. The restoring force on the particle
is proportional to the displacement with a force constant which
for a given particle and beam profile is a linear function of the
laser power. In the experiments detailed below the intensities of
the orthogonal beams were carefully adjusted until the stiffness
of the two traps differed by less than 5\%. The trap stiffness $k$
was typically of the order of 5.1 x 10$^{-6}$Nm$^{-1}$ which
corresponds to a {\it rms} displacement within the trap of about
40nm. The intensity of each beam at the focal plane was estimated
to be of the order of 30mW.

The positions of the two trapped spheres, ${\bm r}_{1}$ and ${\bm
r}_{2}$, were tracked with nanometer resolution by observing the
interference between the transmitted and scattered light in the
back-focal plane of the microscope condenser using a pair of
quadrant detectors \cite{1614}.  Difference voltages from the sum
of the horizontal (X) and vertical (Y) halves of the quadrant
detectors  are linearly proportional to the displacement of the
sphere from the optical axis of the trap. The trajectories, ${\bm
r}_{1}(t)$ and ${\bm r}_{2}(t)$, were measured for a pair of
spheres with mean separations $r = |{\bm r}_{1} - {\bm r}_{2}|$
from between 2.5$\mu$m and 20$\mu$m. For each value of $r$, the
Brownian motion of the two spheres was followed for a total of
420s, at intervals of 50$\mu$sec, to yield 2$^{23}$ (8.4 x
10$^{6}$) samples of the spheres' dynamics.

\section{Brownian motion of confined spheres}
\label{sec:theory}
\subsection{An isolated sphere}
\label{sec:single}

A single hard sphere of radius $a$, moving with a constant
velocity $\bm{{U}}$ through an unbounded fluid of viscosity $\eta$
experiences a  hydrodynamic drag force $\bm{F_{D}}$ in the
direction opposite to motion. In the low Reynolds number limit
\cite{1615} the velocity of the sphere is a linear function of the
force exerted on the particle by the fluid,
\begin{equation}\label{eqn:Stokes}
    \bm{U} = -b_{0}\bm{F_{D}}.
\end{equation}
The constant $b_{0}$ is the mobility of an free particle which, if
there is no slip at the boundary of the particle, is given by
Stokes Law as
\begin{equation}\label{eqn:free mobility}
  b_{0} = \frac{1}{6 \pi \eta a}.
\end{equation}
When the Brownian sphere is confined by a potential, $U (x)$, the
drag force increases. The total force on the particle consists of
a random Gaussian force $f(t)$ together with an additional force
due to the potential field, $-dU / dx$. For a harmonic potential
(of stiffness $k$) in the `long-time' limit, where inertial terms
are negligible, the motion of a confined Brownian sphere is
described by the Langevin equation,
\begin{equation}\label{eqn:Langevin}
  \frac{dx}{dt} = b_{0}[f(t) -k x(t)]
\end{equation}
with a random particle force $f(t)$ which is Gaussian distributed
with the moments,
\begin{eqnarray}\label{eqn:gauss}
  \langle f(t)\rangle  & = & 0 \nonumber  \\
  \langle f(t)f(t')\rangle  & = & 2 b_{0}^{-1} k_{B}T \delta(t-t').
\end{eqnarray}
This Langevin equation is readily solved by standard methods
\cite{Doi-646} and the position autocorrelation function $\langle
x(t)x(0)\rangle $ determined.  Since there is only one
characteristic timescale in the long-time limit, the
autocorrelation decays exponentially
\begin{equation}\label{eqn:harmonic}
    \langle x(t)x(0)\rangle  = \frac{k_{B}T}{k}
    \exp(-t/\tau)
\end{equation}
with a decay time $\tau$ which is physically just the time taken
by a sphere to diffuse a distance $l^{*}$,
\begin{equation}\label{eqn:length}
  l^{*} = \sqrt{\frac{2k_{B}T} {k}},
\end{equation}
where $l^{*}$ is the classical turning point of the confining
potential, or the separation at which the potential energy of the
trapped sphere equals its thermal energy $k_{B}T$.

\subsection{A pair of spheres}

The motion of a pair of harmonically-bound particles differs from
\ref{sec:single} because the hydrodynamic forces couple the motion
of the two spheres together. As one particle moves a flow is
created in the surrounding fluid which drives fluctuations in the
position of a neighbouring second sphere \cite{1615}. In this
section we analyze the correlated motion of a pair of particles
which are coupled by such dynamic forces. We assume, for
simplicity, that (1) the two trapped particles have the same
diameters and are contained within optical traps with identical
force constants; and (2) that there is no potential coupling
between the two spheres.

The hydrodynamic forces acting between equal-sized spheres have
been calculated by a number of authors
\cite{Batchelor-618,1666,Jeffrey-617}. In low-Reynolds number flow
the hydrodynamic interactions between two spheres can be described
by a set of linear relations between the force or torque exerted
on a sphere and the corresponding translational and rotational
velocities. If, as here, the spheres are freely rotating the
applied torque must be zero and one can eliminate the angular
velocities. In this case the linear relation between the forces
and translational velocities defines the mobility tensor
$\bm{b}_{ij}$,
\begin{equation}\label{eqn:tensor}
  \begin{pmatrix}
     \bm{U}_{1} \\
     \bm{U}_{2}
  \end{pmatrix} =
  \begin{pmatrix}
    \bm{b}_{11} & \bm{b}_{12} \\
    \bm{b}_{21} & \bm{b}_{22}
  \end{pmatrix}
  \begin{pmatrix}
     \bm{F}_{1} \\
     \bm{F}_{2}
  \end{pmatrix}
\end{equation}
where the two spheres are labeled $1$ and $2$ and the equivalence
of the two particles implies that $\bm{b}_{11} = \bm{b}_{22}$ and
$\bm{b}_{12} = \bm{b}_{21}$. The spherical symmetry of the Stokes
limit reduces to one of axial symmetry and the mobility tensor
depends crucially on the geometry of the two spheres
\cite{Batchelor-618}. The linearity of the Stokes equation implies
that each of the matrices $\bm{b}_{ij}$ can be decomposed into a
pair of mobility coefficients which describes motion either along
the line of the centres or perpendicular to it,
\begin{equation}
  \bm{b}_{ij}(\bm{r}) = A_{ij}(r) \frac{\bm{r}\bm{r}}{r^{2}} +
  B_{ij} (\bm{1} -\frac{\bm{r}\bm{r}}{r^{2}})
\end{equation}
where the coefficients, $A_{ij}$ and $B_{ij}$, detail the
longitudinal and transverse mobilities, respectively. In the
remainder of this paper we shall confine our discussion to the
longitudinal motion alone. In this case only two mobility
coefficients are needed to quantify the hydrodynamic forces. Each
of these mobilities, $A_{11}$ and $A_{12}$, is itself a function
only of the normalised separation $\rho = r /a $ of the two
spheres.

The mobility coefficient, $A_{12}$, couples the fluctuations along
the line of the centres of the two spheres, chosen here as the
$x$-axis, so that the particle coordinates, $x_{1}$ and $x_{2}$,
are no longer independent. The extent of their correlation is
determined from the solution of the Langevin equation
\begin{equation}\label{eqn:langevin1}
  \begin{pmatrix} \dot{x}_{1} \\ \dot{x}_{2}  \end{pmatrix}
  = \begin{pmatrix}A_{11} & A_{12} \\ A_{12} &  A_{11}
  \end{pmatrix}\begin{pmatrix}f_{1}(t)-kx_{1} \\ f_{2}(t)-kx_{2}
  \end{pmatrix}
\end{equation}
with the random force characterized by the moments
\begin{eqnarray}
  \langle f_{i}(t)\rangle  & = & 0 \nonumber  \\
  \langle f_{i}(t)f_{j}(t')\rangle  & = & 2 (A^{-1})_{ij} k_{B}T \delta(t-t').
\end{eqnarray}
and  $(A^{-1})_{ij}$ as the inverse matrix of $A_{ij}$. Note that
the mobility gradient terms, $k_{B}T/2 \sum_{j} \partial A_{ij}
/
\partial x_{j}$, in the conventional Langevin equation have been
ignored in equation~\ref{eqn:langevin1} because in our experiments
the positional fluctuations are typically two orders of magnitude
smaller than the mean particle separation.

To solve this coupled Langevin equation we introduce the normal
coordinates $X_{i}$,
\begin{equation}\label{eqn:transform}
  X_{i} = \sum_{j} c_{ij} x_{j}
\end{equation}
and choose the coefficients $c_{ij}$ so that the equation of
motion for $X_{i}$ has the following form \cite{Wang-650},
\begin{equation}\label{eqn:mode}
  \frac{ dX_{i}}{dt} = -k \lambda_{i} X_{i} + F_{i}(t)
\end{equation}
with $i$ = (1 or 2). It is readily shown that the matrix $c_{ij}$
consists of the normalised eigenvectors of the mobility matrix
$A_{ij}$ so that the normal modes are
\begin{eqnarray}
\label{eqn:normal} X_{1} & =&  \frac{1}{\sqrt{2}} (x_{1} + x_{2})
\nonumber \\ X_{2} & = & \frac{1}{\sqrt{2}} (x_{1} - x_{2})
\end{eqnarray}
which describe, in turn, a symmetric  collective motion ($X_{1}$)
of the centre of mass of the two spheres and an antisymmetric
relative motion ($X_{2}$) of the two spheres with respect to each
other along the line of their centres. The mobilities
$\lambda_{i}$ of the two modes are the eigenvalues of the matrix
$c_{ij}$,
\begin{eqnarray}\label{eqn:lambda}
 \lambda_{1} & =&  A_{11} + A_{12} \nonumber \\
  \lambda_{2} & = & A_{11} - A_{12}.
\end{eqnarray}
while the $F_{i}$'s are random forces which satisfy
\begin{eqnarray}
  \langle F_{i}(t)\rangle  & = & 0 \nonumber  \\
  \langle F_{i}(t)F_{j}(t')\rangle  & = & 2 \delta_{ij} \lambda_{i} k_{B}T \delta(t-t').
\end{eqnarray}
Since the random forces are independent of each other, motion of
the two normal modes are also independent of each other. The
hydrodynamic term which couples the motion of the two spheres,
$A_{12}$, leads to an asymmetry in the decay times of the normal
modes. The time correlation functions of the normal coordinates
are calculated readily from equation~\ref{eqn:mode} as
\begin{equation}
  \langle X_{i}(t) X_{j}(0) \rangle = \delta_{ij} \frac{k_{B}T}{k} \exp( -t /
  \tau_{i})
\end{equation}
with decay times
\begin{equation}\label{eqn:tau}
  \tau_{i} = \frac{1}{k \lambda_{i}}
\end{equation}

Inverting the coordinate transformation of
equation~\ref{eqn:transform} gives the normalised time correlation
functions of the particle centres
\begin{equation}\label{eqn:norm_correlate}
  h_{ij}(t) =  \frac{\langle x_{i}(t) x_{j}(0) \rangle}{\sqrt{\langle
  x_{i}^{2}\rangle \langle
  x_{j}^{2}\rangle}},
\end{equation}
 as
\begin{eqnarray}\label{eqn:cross}
h_{11}(t) & = & \frac{1}{2}\bigl[
 \exp( -t / \tau_{1}) + \exp( -t/ \tau_{2}) \bigr] \nonumber \\
h_{12}(t) & = & \frac{1}{2}\bigl[
 \exp( -t / \tau_{1}) - \exp( -t/ \tau_{2}) \bigr].
\end{eqnarray}
Inspection reveals that the cross-correlation is very sensitive to
the hydrodynamic coupling between the two spheres. At small times,
$t \rightarrow 0$, $ h_{12}$ records only the time-averaged,
static correlations \cite{Chaikin-1263} which, at thermal
equilibrium, depends only on the interparticle potential. In the
current experiments there is no potential coupling between the two
spheres and so $h_{12}(0) = 0$. At long delay times, $t
\rightarrow \infty$, the hydrodynamic flows which couples the
motion decay to zero so that the positions of the two spheres are
uncorrelated and $h_{12}(t \rightarrow \infty) = 0$. The cross
correlation  is therefore zero both at short and long times.
Equation~\ref{eqn:cross} reveals that the cross correlation will
also be zero at intermediate times unless the two decay times,
$\tau_{1}$ and $\tau_{2}$, differ. From equations~\ref{eqn:tau}
and \ref{eqn:lambda} this difference is a linear function of the
hydrodynamic coupling term $A_{12}$. In the limit where $A_{12}
\ll A_{11}$, which is the case in most physical situations, the
cross-correlation has a minimum at a time which is fixed, to
leading order, by the diagonal mobility $A_{11}$
\begin{equation}\label{eqn:tmin}
  t^{*} = \frac{1}{k A_{11}} \biggl \{ 1 + \frac{1}{3}
  \biggl(\frac{A_{12}}{A_{11}}\biggr )^{2} + O \biggl[\biggl(\frac{A_{12}}{A_{11}}\biggr )^{4}\biggr ]\biggr \}
\end{equation}
while the depth of the minimum is determined by the ratio of the
off-diagonal and diagonal mobilities
\begin{equation}\label{eqn:min}
  h_{12}(t^{*}) = - \frac{1}{e} \biggl \{
  \frac{A_{12}}{A_{11}}  + O \biggl[\biggl(\frac{A_{12}}{A_{11}}\biggr )^{3}\biggr ]\biggl \}.
\end{equation}

\section{Results}
\label{sec:results}

\begin{figure}
   \centering
   \includegraphics[width=5in]{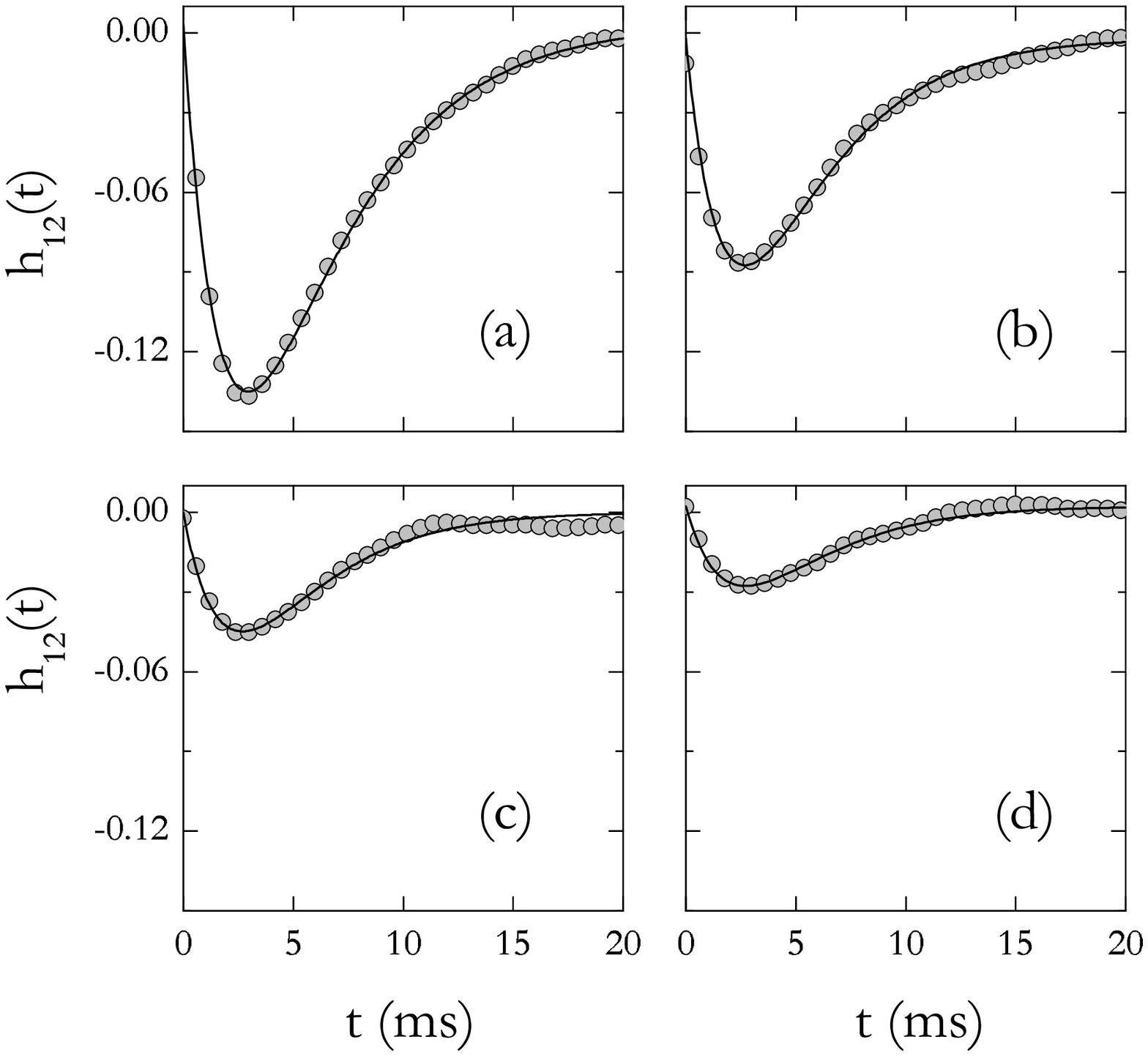}
   \caption{Cross correlation functions for 1.30
   $\mu$m diameter PMMA spheres as a function of delay time $t$ and at four centre-to-centre separation $r$.
   (a) $r = 2.47\mu$m, (b) $r = 4.19\mu$m, (c) $r = 7.44\mu$m,  and (d) $r = 11.47\mu$m. The motion is measured
   parallel to the separation vector. The solid line shows a fit to equation~\ref{eqn:cross}. For clarity only every
   twelve data point is plotted.}
   \label{fig:correl}
   \end{figure}

The normalised longitudinal position cross correlation,
$h_{12}(t)$, was measured for a pair of trapped spheres over a
wide range of separations (2.5 $\mu$m $< r < 20$ $\mu$m).
Figure~\ref{fig:correl} shows typical data for four sphere
separations.  Three features in the experimental data are
striking. First, the data show, rather surprisingly, that
hydrodynamic interactions cause the two particles to be {\it
anti-correlated} at intermediate times; second, the time at which
the two particles are most strongly anti-correlated, the time
$t^{*}$ at the minimum of $h_{12}$ , does not vary with the sphere
separation; while conversely, the strength of the anti-correlation
increases markedly as the separation $r$ reduces.

To interpret these observations the decay times $\tau_{1}$ and
$\tau_{2}$ of the symmetric and anti-symmetric normal modes were
extracted from a least squares fit of the data to
equation~\ref{eqn:cross}. The quality of the resulting fit may be
gauged, for the four separation depicted in
figure~\ref{fig:correl}, by studying the solid curves, which are
seen to accurately reproduce the measured data. From the
experimentally determined decay times and trap stiffness the
elements of the mobility tensor may be estimated as,
\begin{eqnarray}\label{eqn:mobility}
A_{11} &=& \frac{1}{2k} \biggl ( \frac{1}{\tau_{1}} +
\frac{1}{\tau_{2}} \biggr ) \nonumber \\ A_{12} &=& \frac{1}{2k}
\biggl ( \frac{1}{\tau_{1}} - \frac{1}{\tau_{2}} \biggr ).
\end{eqnarray}

\begin{figure}
   \centering
   \includegraphics[width=5in]{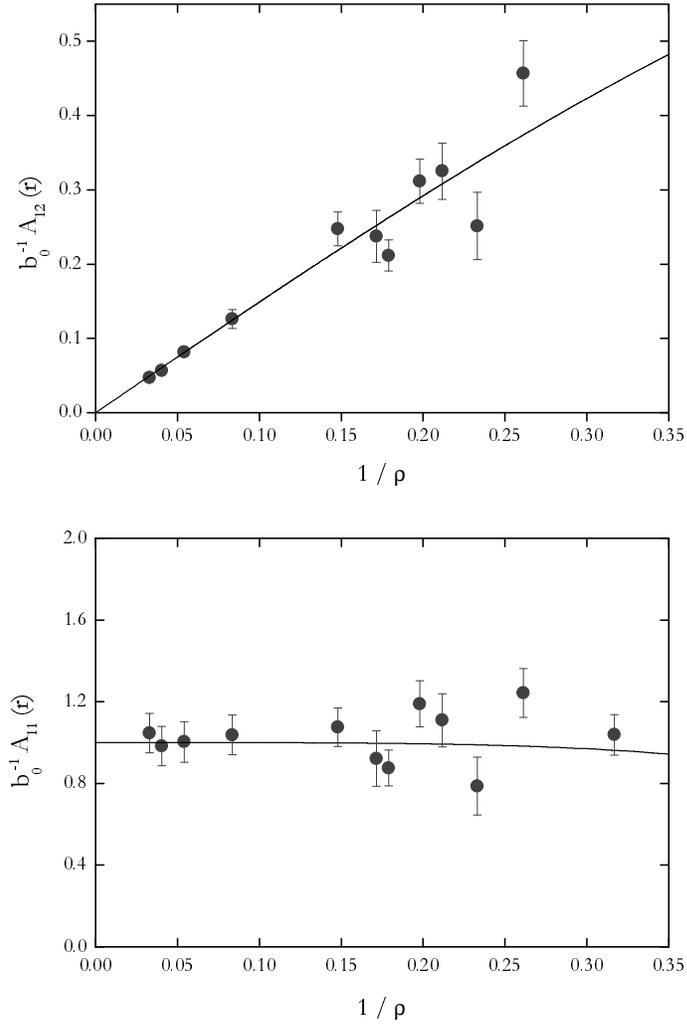}
   \caption{The experimentally-determined mobility coefficients,
   $b_{0}^{-1}A_{ij}$, for motion along
   the line of centres as a
   function of the inverse centre-to-centre separation $\rho = r/ a$ . Here $b_{0}$ is the mobility of the PMMA particle of radius $a$. Solid lines show the
   predictions of low-Reynolds number hydrodynamic calculations
   for the case of two interacting solid
   spheres \cite{Batchelor-618}, with no adjustable parameters.}
    \label{fig:mobility}
   \end{figure}

The experimentally-determined scaled mobility elements,
$b_{0}^{-1} A_{ij}(r)$, are plotted in figure~\ref{fig:mobility}
as a function of the dimensionless separation of the two spheres,
$\rho = r /a$. The two mobilities show a strikingly different
dependence on the sphere separation. While the diagonal mobility
is largely unaffected by the sphere separation the off-diagonal
term scales approximately inversely with $\rho$. These trends are,
of course, consistent with the observations made above that the
position, $t = t^{*}$, of the minimum in the correlation function
does not shift with separation (see equation~\ref{eqn:tmin}) while
the depth increases with reducing separation
(equation~\ref{eqn:min}).

The experimental values for the mobilities may be compared with
theoretical predictions for the hydrodynamic coupling of two hard
spheres. Batchelor \cite{Batchelor-618}, for instance, has given
the following expressions for the longitudinal mobilities,
\begin{eqnarray}
b_{0}^{-1} A_{11} & = &  1 - \frac{15}{4 \rho^{4}} + O(\rho^{-6})
\nonumber \\ b_{0}^{-1}  A_{12} & = & \frac{3}{2\rho} -
\frac{1}{\rho^{3}} + O(\rho^{-7})
\end{eqnarray}
which are exact in the limit of large centre-to-centre separation
$\rho$, as has been confirmed by \cite{1666}. The solid curves in
figure~\ref{fig:mobility} show the predictions of the Batchelor
\cite{Batchelor-618} theory for the longitudinal coupling. As is
clear from this figure, the measured mobilities agree very well
with the theoretical predictions over the entire experimentally
accessible range of separations.

\begin{figure}
  \centering
   \includegraphics[width=5in]{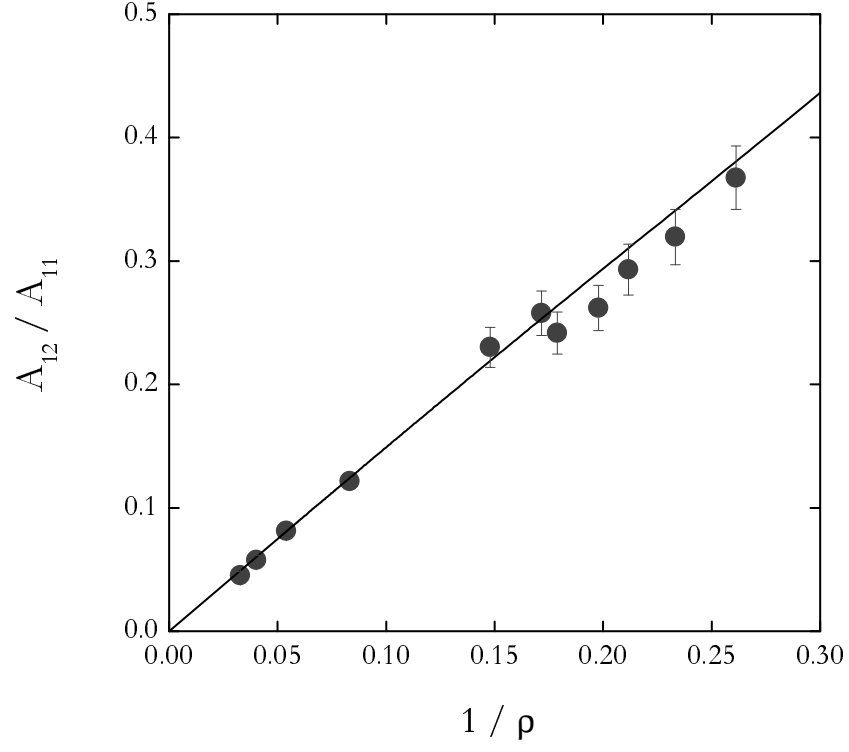}
   \caption{The experimental longitudinal mobility ratio $A_{12}/A_{11}$ as a function of the
   inverse sphere separation $\rho$, in units of the sphere radius. The solid line shows the
   predictions of \cite{Batchelor-618}.}
   \label{fig:mobility ratio}
   \end{figure}

The deviations between theory and experiment evident  in
figure~\ref{fig:mobility} are due largely to the experimental
difficulty measuring the force constant $k$. This is seen in
figure~\ref{fig:mobility ratio} where the experimentally
determined mobility ratio, $A_{12} / A_{11}$, which from
equation~\ref{eqn:mobility} does not require any knowledge of $k$,
is plotted as a function of the inverse separation $1 / \rho$. The
almost quantitative agreement seen, {\it with no adjustable
parameters}, between the data and theory confirms the accuracy of
Batchelor's theoretical description of pair hydrodynamics. In
addition the close agreement between experiment and theory
suggests that, at least for the range of distances explored in the
current experiments, the hydrodynamics forces between
polymer-coated and uncoated spheres are very comparable.

\section{Discussion}

We have presented a detailed experimental study of hydrodynamic
coupling between an isolated
 pair of polymer-coated hard-sphere colloids.  We find near quantitative agreement
with Low Reynolds-number predictions for the hydrodynamic coupling
between a pair of spheres. Surprisingly, we observe a strong {\it
anti-correlation} in the positions of the two coupled spheres at
intermediate times. At first sight this result looks
counter-intuitive since one might expect naively a symmetric
correlation between spheres. However the effect is a dynamic
time-dependent phenomenon. The origin of which may be understood
from the normal modes of the system. The motion of two spheres,
along the line of their centres, decouples when analyzed in terms
of a symmetric collective mode and an anti-symmetric relative
mode. The mobilities of these independent modes are, from the
asymptotic expressions of Batchelor \cite{Batchelor-618} and
equation~\ref{eqn:lambda},
\begin{eqnarray}
\label{eqn:mobile} \lambda_{1} & = & b_{0} \left \{ 1 + \frac{3}{2
\rho} - \frac{1}{\rho^{3}} - \frac{15}{4 \rho^{4}} + O(\rho^{-6})
\right \} \nonumber \\ \lambda_{2} & = & b_{0} \left \{ 1 -
\frac{3}{2 \rho} + \frac{1}{\rho^{3}} - \frac{15}{4 \rho^{4}} +
O(\rho^{-6}) \right \},
\end{eqnarray}
where $\rho$ is the dimensionless centre-to-centre separation,
$\rho = r /a$. Examination of the leading terms in this equation
reveals that the mobility $\lambda_{1}$ of the symmetric mode is
enhanced and the anti-symmetric mode reduced when compared with an
isolated particle. The reduction in mobility of the anti-symmetric
mode reflects the difficulty of squeezing fluid out of or into the
narrow gap between  two approaching spheres while the increased
mobility for the symmetric mode is caused by the tendency for the
fluid flow generated by one sphere to entrain a neighbouring
sphere.

The asymmetry in the mobilities of the normal modes, seen in
equation~\ref{eqn:mobile}, causes the decay times for thermal
fluctuation in the two modes to differ. When the two spheres are
close together, the mobility of the symmetric mode is enhanced
compared with the anti-symmetric mode. As a consequence, symmetric
fluctuations decay more rapidly than their anti-symmetric
counterparts. At $t = 0$ the proportions of thermally-excited
symmetric and anti-symmetric fluctuations are equal since the
positions of the two spheres are uncorrelated. With increasing
time the amplitudes of both fluctuations decay. However the
anti-symmetric fluctuations decay at a {\it slower} rate than the
symmetric fluctuations  so that the cross correlation develops a
pronounced anti-correlation. The anti-correlation is however
dynamical since over long time all fluctuations decay and the
spheres become again uncorrelated.

In summary, we have shown that 2-particle cross-correlation
spectroscopy (TCS) is a promising new technique for the
quantitative determination of hydrodynamic interactions. TCS
experiments are very flexible; the particle size, separation,
potential interactions and indeed the dispersion medium can all be
changed independently of each other. Variations of the methods
described in this paper could, for instance, be used to follow the
time course of collective fluctuations in a dense (host) complex
fluid from the real-space trajectories of inserted {\it probe}
colloidal particles. Currently, we are utilising TCS to
study the many-body hydrodynamic interactions in concentrated
particulate suspensions. Two probe PMMA particle are trapped
within an index-matched silica suspension of volume fraction
$\phi$. The resulting fluctuations in the trajectories of the two
probe PMMA particles are used to determine the effective
pair-mobility tensor in the {\it suspension}, as a function of
 particle separation and $\phi$. These measurements promise to
provide new and detailed experimental information on the spatial
and temporal development of hydrodynamic interactions in
concentrated suspensions.

\begin{acknowledgements}
This work was supported by a grant from the UK Engineering and
Physical Science Research Council (No. GR/L37533). We thank
Professor R.M. Simmons, Dr R.B. Jones and Dr J.S. van Duijneveldt,
for useful discussions and comments. We also thank Andrew Campbell
for the preparation of the colloidal particles used.
\end{acknowledgements}
%

%
%
  \newpage

\label{lastpage}

\end{document}